\documentclass[sigconf,screen]{acmart} %
\usepackage{graphicx} %
\usepackage{natbib}  %
\usepackage{booktabs} %

\setcopyright{none}
\setcopyright{none}
\renewcommand\footnotetextcopyrightpermission[1]{}

\acmConference[WebSci '25]{17th ACM Web Science Conference 2025}{May 20--24,
  2025}{New Brunswick, NJ, USA}

\setcopyright{acmlicensed}
\copyrightyear{2025}
\acmYear{2025}
\acmDOI{XXXXXXX.XXXXXXX}

\usepackage{comment}
\usepackage{tabularx}
\usepackage{colortbl}
\usepackage{indentfirst}
\usepackage{xspace}
\usepackage{graphicx}
\usepackage{amsmath} 
\newcommand{\descr}[1]{\vspace{0.2cm}\noindent\textbf{#1}}

\usepackage[flushmargin,hang,stable]{footmisc}

\renewcommand{\footnoterule}{
	\kern -3pt
	\hrule width 1in
	\kern 2pt
}

\definecolor{yellowish}{RGB}{168, 115, 30}
\definecolor{headerBlue}{RGB}{111, 168, 220} 
\definecolor{subheaderBlue}{RGB}{181, 210, 237} 
\definecolor{headerBrown}{RGB}{220 168 111} 
\definecolor{subheaderBrown}{RGB}{237 219 181} 
\definecolor{headerPink}{RGB}{237 182 181} 
\definecolor{subheaderPink}{RGB}{246 218 217} 
\definecolor{headerYellow}{RGB}{255, 217, 102} 
\definecolor{subheaderYellow}{RGB}{255, 242, 204} 

\newcommand{\algo}{\textit{VIKI}\xspace}

\newcommand{\personalInterestsDef}{fun hobbies or activities that people like to do outside of work and school, like playing games or reading\xspace}
\newcommand{\professionalInterestsDef}{areas of work that people are involved in\xspace}
\newcommand{\popularDef}{Github authors that have at least one repository with at least five stars and five forks\xspace}

\newcommand{\datasetPopular}{{\it D\textsubscript{Pop}}\xspace}
\newcommand{\datasetMalware}{{\it D\textsubscript{Mal}}\xspace}
\newcommand{\datasetLarge}{{\it D\textsubscript{Lrg}}\xspace}
\newcommand{\datasetGroundtruth}{{\it D\textsubscript{Val}}\xspace}

\newcommand{\threshStdev}{{\it T\textsubscript{stdev}}\xspace}
\newcommand{\threshPosts}{{\it T\textsubscript{posts}}\xspace}

\definecolor{dkgreen}{rgb}{0,0.6,0}

\title{\algo: Systematic Cross-Platform Profile Inference\\of Online Users$^*$}

\usepackage{balance} 

\begin{document}
\sloppy

\author{Ben Treves, Emiliano De Cristofaro, Yue Dong, Michalis Faloutsos}
\email{{btrev003,emiliand,yue.dong,michalisf}@ucr.edu}
\affiliation{%
    \institution{UC Riverside}
    \city{Riverside}
    \state{California}
    \country{USA}
}

\setcounter{secnumdepth}{3} %

\pagestyle{plain}

\begin{abstract}
What can we learn about online users by comparing their profiles across different platforms?
We use the term profile to represent displayed personality traits, interests, and behavioral patterns (e.g., offensiveness).
We also use the term {\it displayed personas} to refer to the personas that users manifest on a platform.
Though individuals have a single real persona, it is not difficult to imagine that people can behave differently in different ``contexts'' as it happens in real life (e.g., behavior in office, bar, football game).
The vast majority of previous studies have focused on profiling users on a single platform.
Here, we propose \algo, a systematic methodology for extracting and integrating the displayed personas of users across different social platforms.
First, we extract multiple types of information, including displayed personality traits, interests, and offensiveness.
Second, we evaluate, combine, and introduce methods to summarize and visualize cross-platform profiles.
Finally, we evaluate \algo on a dataset that spans three platforms -- GitHub, LinkedIn, and X.
Our experiments show that displayed personas change significantly across platforms, with over 78\% of users exhibiting a significant change.
For instance, we find that neuroticism exhibits the largest absolute change. We also identify significant correlations between offensive behavior and displayed personality traits.
Overall, we consider \algo as an essential building block for systematic and nuanced profiling of users across platforms. 

\end{abstract}

\maketitle %

\renewcommand*{\thefootnote}{\fnsymbol{footnote}}
\footnotetext{$^*$Published in the Proceedings of the 17th ACM Web Science Conference (WebSci 2025). Please cite the WebSci version.}
\renewcommand*{\thefootnote}{\arabic{footnote}}

\pagestyle{plain}

\section{Introduction}
\label{sec:intro}

How does the displayed persona of a user change across platforms?
Social media provides individuals with a plethora of platforms to express themselves to the public~\cite{golbeck2011predicting}.
In a way, it enables them to create representations of themselves and their personalities~\cite{baym2015personal}.
These representations can vary across the different platforms they engage in, as they often serve different social purposes, involve different circles of friends/connections, and are generally associated with different kinds of content.
Understanding users and their personality from social media is a highly sought after capability for both research and practical applications. 
Social sciences and psychology could benefit by  
understanding how people respond to an event and by associating to different personality groups. 
\begin{figure}[t]
\centering
\includegraphics[width=0.95\linewidth]{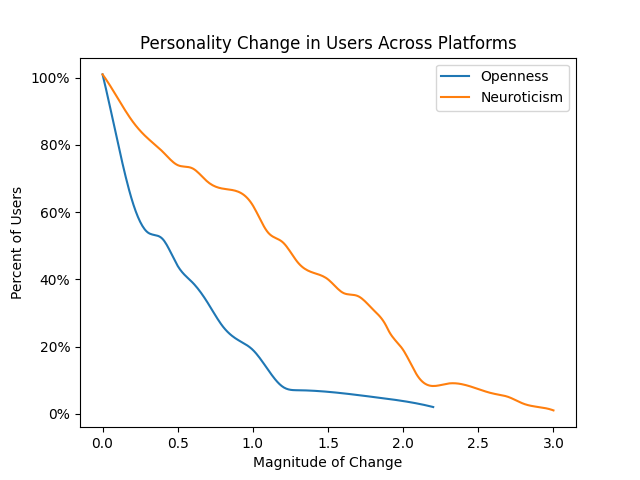}
    \caption{
    Some users display significantly different personalities in different platforms.
    We show the change in the openness and neuroticism traits (on a 1-5 scale) of users across LinkedIn and X using a Complementary Cumulative Distribution Function (CCDF).
	Over 60\% of users display a change in neuroticism by at least one point and 20\% by at least two points between the two platforms. 
    The distribution of change is substantial but  less pronounced for openness.
    }
    \label{fig:personalityChangeCdf}
\end{figure}

We provide some definitions that we use in this work. 
Here, we use the term {\bf persona} or {\bf profile} of a user to refer to both: (a) the {\bf personality} traits, such as extroversion, and (b) the preferences and interests  of a user. 
We use the term {\bf displayed persona} to refer to the persona that a user manifests on a platform, which could be different from the user's true persona. 
Intuitively, one can expect that the displayed personas of a user on different platforms can vary the same way people's behavior changes depending on context (e.g. swearing at a sports arena is tolerated unlike at a business presentation). 
In fact, this is one of the questions that we want to answer here. 
What can we learn about a user
from their displayed personas across multiple platforms?
This is the question that motivates this work.
In fact, we decompose this question into the following related questions:
(a) how can we compare the profile of displayed personas from different platforms,
(b) how can we identify the effect of a platform on the displayed personality of its users, and
(c) are there any correlations between the displayed profile characteristics?
Answering the above questions can provide a critical capability:
we can develop a {\bf cross-platform profile} for a user by critically synthesizing their displayed personas.

Most previous work has focused on user profiling on a single platform or does not address the problem as framed above.  We can place related work into the following groups.
First, we have single-platform studies, which use a plethora of methods to extract the displayed persona of a user from a platforms on~\cite{carducci2018twitpersonality,christian2021text}, 
Second, some studies consider two platforms, but they usually focus on: (a) user disambiguation (matching user identities) \cite{masud2023geekman} ~\cite{goga2015reliability}, or (b) answering niche and targeted questions, such as correlating personality with a certain behavior~\cite{samani2018cross} or identifying preferences, such as the user's taste in music ~\cite{gu2018cross}).

As our key contribution, we present \algo, a novel methodology for extracting, comparing, and integrating the displayed personas of users across platforms.
\algo extracts multiple attributes using a combination of tools as well as 
Large Language Models (LLMs) on user-generated text.
The goal of our work is twofold.
First, we want to create a cross-platform profile by integrating all their displayed personas and second, we want to quantify and highlight the discrepancies in these personas.
\algo elicits the displayed personality traits %
based on the ``Big-Five'' model  (a.k.a. OCEAN), namely, openness, conscientiousness, extraversion, agreeableness, and neuroticism~\cite{goldberg1990alternative}.\footnote{This is the personality that users display on a given platform and is not necessarily equal to their ``true'' individual personality; hence, we stress that our study only makes inferences and considerations on {\em displayed} personalities.}
The pipeline of our method
consists of the following steps:
(a) we infer the personality on each platform,
(b) we extract the professional and personal interests,
(c) we compare the difference of the displayed personalities of one user across platforms,
and (d) we quantify the effect that a platform seems to have on the displayed persona of its users.
Finally, we use clustering to identify users: (a) with similar personalities and (b) with similar changes in personality across platforms.
We evaluate our approach on a dataset of 1.2K online tech-oriented users as we explain below. 
First, we want to state that this task required us to be creative, since identifying this dataset entails a large number of users with accounts on the same platforms (for comparison purposes), which also had to be publicly accessible.
Our solution was to start from GitHub users who often self-report their social media handles.
In our set of users, LinkedIn and X (formerly Twitter) were the most widely shared social media handles. 
Serendipitously, these two social media have fairly distinct intent and use:
X is typically associated with general social use, while LinkedIn is a professional-oriented networking platform.

We provide a sample of the key results from our study below.

\begin{itemize}
\item Most users (78\%) change by at least one point and one in four (25\%+) by at least two points in at least one personality trait across platforms.
A few (4\%) exhibit a personality change of at least one point in \textit{all five} OCEAN traits.
On average, all five OCEAN traits change by at least half of a point across platforms.
Neuroticism yields the largest absolute change (1.2 points), while openness the smallest (0.5 points), as visualized %
in Figure~\ref{fig:personalityChangeCdf}.

\item We find a significant directional change in OCEAN traits across platforms, with users in our dataset displaying 1.1 higher neuroticism and 0.4 lower conscientiousness on X, compared to LinkedIn, on average.
Two-sample Kolmogorov– Smirnov (KS) tests~\cite{smirnov1939estimation} confirm statistically significant differences across LinkedIn and X for the CDFs of both neuroticism and conscientiousness with, respectively, $D=0.63$ ($p < 0.001$) and $D=0.46$ ($p< 0.001$). %

\item Neurotic users in our dataset are also more likely to be offensive.
We find moderate correlations between neuroticism and offensive behavior for both malicious and popular GitHub users ($R=0.37$ and $0.25$ with p-values of 0.00 and 0.01, respectively).
\end{itemize}

Naturally, these observations are specific to the datasets at hand and should not be used to provide wide generalizations; however, we see them as a promising indication that cross-platform user analysis is worthwhile.
In other words, single-platform profiling of these users could yield notably different displayed personalities for each platform, while cross-platform profiling paints a more comprehensive and nuanced picture.
Arguably, our work paves the way for practitioners and researchers to compare and contrast users' online personas as a means of gaining insight into how individuals adapt themselves to different online environments and the varying influence that platforms have on their users.
\begin{figure*}[t]
\centering
\includegraphics[width=0.95\textwidth]{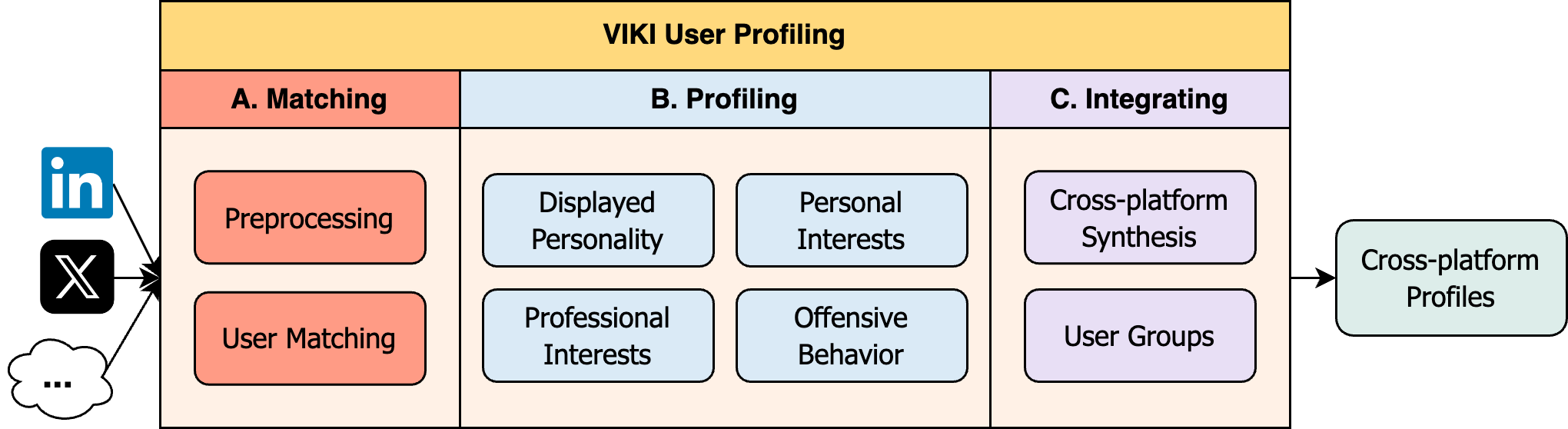}
    \caption{The \algo methodology for cross-platform user profiling.}
    \label{fig:method_flowchart}
\end{figure*}

\section{The \algo Methodology}
\label{sec:methodology}

In this section, we present \algo, our novel cross-platform user profiling methodology.
As outlined in Figure~\ref{fig:method_flowchart}, \algo includes three key components geared for:
1) matching users across social platforms,
2) profiling users on each platform, and
3) integrating single-platform profiles into a cross-platform user profile.

\subsection{Matching}

Our cross-platform user profiling requires matching user identities  across platforms platforms.
We present this part for completeness, but we do not claim to have an algorithmic contribution at this point, though this can change in the future.
In fact, this step can be addressed by either finding readily available matched identities or by developing an approach to match identities.

One approach is to identify sets of users who explicitly report and link their user handles on different platforms.
We can consider two "filtering" steps.
First, we want to select users that have presence in the same social platforms to have consistency in the study.
Second, in the case of a large set of users
we can randomly select a sample of users, especially if further processing (e.g. crawling user profiles) is computationally intensive.
Finally, we use both automated and manual efforts to verify that the linked accounts are active.
We then deploy methods to collect a sufficient number of activities in each platform (e.g. personal statement, posts, comments etc).

Formally, let $U$ be the set of users with external accounts, %
$P$ the set of social media platforms, and $A(u)$ the set of accounts linked by user $u \in U$. 
We define $S \subset U$ as users who self-report accounts across multiple platforms, $V(u)$ as a function verifying the activity of $u$'s linked accounts, and $C(u, p)$ as the function collecting recent posts by $u$ on $p \in P$.
The subset of users $S$ is determined as:
\[
S = \{ u \in U \mid A(u) \cap P \neq \emptyset \}
\]
The users in $S$ are sampled, and for each user $u \in S$, the  verification and content collection are defined as follows:
\[
V(u) = \text{True} \text{ if } \forall p \in A(u), p \in P \text{ and the account is active.}
\]
Once $V(u)$ is verified, the most recent posts are collected as:
\[
C(u, p) = \text{Recent posts of } u \text{ on platform } p.
\]
The overall process can be summarized as:
\begin{equation}
\{ C(u, p) \mid u \in S, V(u) = \text{True}, p \in A(u) \cap P \}  
\end{equation}

\subsection{Profiling}\label{sec:profiling}
Our approach profiles users on each active platform using four aspects: 
1) displayed personality,
2) professional interests,
3) personal interests, and
4) offensive behavior.

Formally, for each user $u \in S$ and platform $p \in A(u) \cap P$, we define a profiling function $F(u, p)$ that computes the profile of user $u$ on platform $p$. This profile is a combination of the four attributes as described below.

\subsubsection{Displayed Personality.}
We model a user’s displayed personality using the ``Big Five'' personality traits, namely, openness, conscientiousness, extraversion, agreeableness, and neuroticism, aka the OCEAN model~\cite{goldberg1990alternative}.
This choice is grounded in psychology, where tests like the International Personality Item Pool (IPIP)~\cite{goldberg1999broad} are commonly used to assess an individual's personality. 
However, these methods are not scalable in an online context, as they require the active participation and compensation of a large number of users. 
Fortunately, prior research has shown that a user's displayed personality can be inferred from their social media posts, especially with the advancements in LLMs~\cite{peters2024large}.
This yields the personality that a user {\em displays} on a given platform and is not necessarily equal to their ``true'' individual personality.

Let $\text{Posts}(u, p)$ represent the set of recent posts by user $u$ on platform $p$. We use an LLM-based method (Google Gemini LLM~\cite{team2023gemini}) to estimate personality traits by feeding $\text{Posts}(u, p)$ into the model. 
Let $T_i(u, p)$ represent the score of the $i$-th personality trait for user $u$ on platform $p$, where $i \in \{1, 2, 3, 4, 5\}$ corresponds to openness, conscientiousness, extraversion, agreeableness, and neuroticism, respectively. The trait scores are obtained by averaging $\kappa$ independent runs of the model for each trait:
\begin{equation}
    T_i(u, p) = \frac{1}{k} \sum_{j=1}^{\kappa} \text{LLM}(\text{Posts}(u, p), i)
\end{equation}

We exclude users with a standard deviation of trait scores higher than a threshold $\sigma_0$, i.e., if:
\[
\text{std}(\{T_i(u, p)\}) > \sigma_0 \quad \text{for any } i
\]
We set $ \sigma_0 = 0.6$, indicating high variability and potentially less accurate profiling.
Naturally, practitioners can tune this parameter to adjust it to their specific needs.

The final personality profile of user $u$ on platform $p$ is represented as a five-dimensional vector:
\[
\mathbf{T}(u, p) = [T_1(u, p), T_2(u, p), T_3(u, p), T_4(u, p), T_5(u, p)]
\]

\subsubsection{Professional Interests.} 
As a user interacts on an online platform, they may potentially discuss some of their professional interests, i.e., \professionalInterestsDef, on some of them. %
\algo aims to collect them as they can be an integral part of a user's profile.
Moreover, we are once again interested in determining whether they are consistent, or at all similar, across platforms. 

\algo detects a user's interests by feeding their most recent posts to Google Cloud's NLP content classification module, which returns a set of content categories that apply to the given text. These categories are hierarchical, with 27 top-level categories and 1,091 leaf-level categories.

Formally, let $\mathcal{C}(\text{Posts}(u, p))$ be the set of content categories returned by an NLP-based classifier. The categories are hierarchical, with $l_1$ top-level categories and $l_2$ leaf-level categories. Let $\mathcal{P}$ denote the set of categories identified as professional interests.
We define the professional interests of user $u$ on platform $p$ as:
\begin{equation}
    \mathcal{I}_{\text{prof}}(u, p) = \mathcal{C}(\text{Posts}(u, p)) \cap \mathcal{P}
\end{equation}

The granularity of the categories can be adjusted by choosing different levels of the hierarchy.

\subsubsection{Personal Interests.}
Similarly, we extract a user's personal interests by identifying content categories unrelated to professional interests, %
e.g., \personalInterestsDef.
Let $\mathcal{Q}$ represent the complement of the set of professional-related categories $\mathcal{P}$, such that $\mathcal{Q} = \mathcal{C}(\text{Posts}(u, p)) \setminus \mathcal{P}$.

The personal interests of user $u$ on platform $p$ are defined as:
\begin{equation}
\mathcal{I}_{\text{pers}}(u, p) = \mathcal{C}(\text{Posts}(u, p)) \cap \mathcal{Q}
\end{equation}
This way, we extract the set of a user's personal interests on a platform, with an adjustable granularity for each interest.

\subsubsection{Offensive Behavior.}
Finally, we focus on users' offensive behavior, which can be an important part of their persona, too.
This will also allow us to assess whether users act offensively and consistently across platforms.
To do so, we run PerspectiveAPI~\cite{lees2022new} on the user's recent posts.
PerspectiveAPI is an online automated model for detecting offensive language. For each post, we extract the score for each of the six production attributes: toxicity, severe toxicity, identity attack, insult, profanity, and threat. Let $\text{O}_a(u, p)$ denote the offensiveness score for attribute $a$ (e.g., toxicity, insult, profanity) of user $u$ on platform $p$, and let $A$ represent the set of offensive attributes (toxicity, severe toxicity, identity attack, insult, profanity, threat).

We classify user $u$ as offensive on platform $p$ if any of their posts exceed a threshold $\tau$ for any attribute $a \in A$:
\begin{equation}
\text{Offensive}(u, p) = 
\begin{cases}
1 & \text{if } \exists a \in A, \text{O}_a(u, p) \geq \tau \\
0 & \text{otherwise}
\end{cases}
\end{equation}
We use an offensiveness threshold such that if any post from a user has a score greater than or equal to the threshold for any of the attributes, the user is labeled as offensive. By default, we recommend an offensiveness threshold value of 0.8, which surpasses baseline thresholds for offensiveness classification based on a single attribute~\cite{hua2020towards}. However, this threshold can be adjusted depending on the specific use case.

\subsubsection{Overall Profile.}
The overall profile $F(u, p)$ for user $u$ on platform $p$ combines the displayed personality $\mathbf{T}(u, p)$, professional interests $\mathcal{I}_{\text{prof}}(u, p)$, personal interests $\mathcal{I}_{\text{pers}}(u, p)$, and offensive behavior $\text{Offensive}(u, p)$:
\[
F(u, p) = \{\mathbf{T}(u, p), \mathcal{I}_{\text{prof}}(u, p), \mathcal{I}_{\text{pers}}(u, p), \text{Offensive}(u, p)\}
\]

This profiling process allows us to analyze users across multiple platforms and compare their behavior and interests in a structured, formal manner.

\subsection{Integration}\label{sec:integrate}

\descr{User Clustering.} To capture trends and similarities between users, \algo also includes a clustering module to group ``similar'' users.
For instance, users could be grouped based on the similarity of their single-platform profiles or interests, the similarity of their cross-platform change in profiles or interests, etc.
Once again, this component can be instantiated in different ways; for simplicity, we only discuss clustering users based on the similarity of their displayed personality traits across a single platform.

Let $\mathbf{T}(u, p)$ denote the five-dimensional vector representing the OCEAN personality traits of user $u$ on platform $p$.
We aim to group users based on their personality vectors.
We use $k$-means clustering~\cite{macqueen1967some} to group users with similar displayed personalities. 
Each user is represented by a five-dimensional vector $\mathbf{T}(u) = [T_1(u), T_2(u), T_3(u), T_4(u), T_5(u)]$ where $T_i(u)$ is the averaged personality trait score of user $u$ across platforms:
\[
T_i(u) = \frac{1}{|A(u) \cap P|} \sum_{p \in A(u) \cap P} T_i(u, p)
\]

Thus, each user $u$ is represented by a vector $\mathbf{T}(u)$, and we use $k$-means clustering to partition the users into $k$ clusters based on this vector.

To find the optimal number of clusters $k$, we use the silhouette score method~\cite{shahapure2020cluster}.
Specifically, we test different values of $k$ within the range $k \in [2, 20]$ and select the value of $k$ that maximizes the average silhouette score:
\[
k_{\text{optimal}} = \arg\max_k \frac{1}{N} \sum_{i=1}^{N} \text{silhouette}(i, k)
\]
where $\text{silhouette}(i, k)$ is the score of the $i$-th user for a clustering with $k$ clusters, and $N$ is the total number of users.
Each $k$-means clustering is repeated ten times to account for randomness, and the clustering with the highest average silhouette score is selected.

\descr{Cross-Platform Synthesis.} The final step of \algo is integrating the single-platform profiles into a unified cross-platform profile.
This may or may not include the results of the clustering module, whereby the similarity of users and the groups they belong to can be part of their cross-platform profile.
To ease presentation, we again only consider users' OCEAN traits, although the synthesis can be performed over their interests, offensiveness, etc.

Let $F(u, p)$ denote the profile of user $u$ on platform $p$ as defined earlier.
The goal is to combine these profiles across platforms $p \in A(u) \cap P$ to form a comprehensive profile of user $u$.
Formally, let $\mathcal{F}(u)$ represent the cross-platform profile of user $u$.
This is defined as the aggregation of their individual platform profiles:
\begin{equation}
 \mathcal{F}(u) = \{ F(u, p) \mid p \in A(u) \cap P \}   
\end{equation}

This synthesis allows for a comparative analysis of the different aspects of a user's persona across platforms, enabling us to evaluate whether the traits, interests, and behaviors are consistent or vary significantly between platforms.

\section{Experimental Evaluation}
\label{sec:evaluation}

\noindent
We present the experimental evaluation of our approach.

\subsection{Personality Inference Validation}

\begin{table}[t]
    \centering %
    		\setlength{\tabcolsep}{5pt}
\begin{tabular}{lrrrrr}
\toprule
\textbf{Error} & \textbf{Open} & \textbf{Cons} & \textbf{Extr} & \textbf{Agre} & \textbf{Neur} \\
\midrule
\textbf{RMSE} & 1.266 & 1.319 & 1.009 & 0.894 & 1.015 \\
\textbf{MSE} & 1.603 & 1.739 & 1.017 & 0.799 & 1.029 \\
\textbf{MAE} & 1.025 & 1.070 & 0.811 & 0.711 & 0.804 \\
        \bottomrule
    \end{tabular}
\caption{Average per-trait errors of \algo's LLM-based personality inference on the {\sl myPersonality} dataset (\datasetGroundtruth).} %
    \label{table:evalResults}
\end{table}

First of all, we set out to validate the viability of \algo's LLM-based personality inference module.
To do so, we apply it on the {\sl myPersonality} dataset~\cite{celli2013workshop}, which we denote as \datasetGroundtruth.
This dataset includes 9,917 posts by 250 users, with each post labeled for the user's true OCEAN personality traits.
In Table~\ref{table:evalResults}, we report the average error values for each trait, using the Root Mean Squared Error (RMSE), as done in prior work~\cite{quercia2011our}, as well as the Mean Squared Error (MSE) and Mean Average Error (MAE).

We apply \algo on \datasetGroundtruth following the methodology presented in Section~\ref{sec:profiling}, i.e., we ask the LLM to infer the five OCEAN personality traits of each user ten times and take the averages.
We use the standard deviation across these inferences and the number of posts per user as thresholds for filtering out \datasetGroundtruth users, which we denote, respectively, as \threshStdev and \threshPosts.
We evaluate the following four thresholds per user:
{\em i)} no thresholds,
{\em ii)} $\threshStdev <= 0.6$,
{\em iii)} $\threshPosts > 1$,
{\em iv)} both $\threshStdev <= 0.6$ and $\threshPosts > 1$.
Ultimately, we find that {\em iv)} yields the lowest RMSE values in every trait, i.e.,  \algo yields more accurate personality inferences when filtering users with little to no generated content.

We conclude that \algo's personality inference module 
is robust, as it provides consistent results and
its accuracy is on par with that of prior methods~\cite{quercia2011our}.

\subsection{Datasets}
\label{sec:dataset}

\begin{table}[t]
\centering
    \begin{tabular}{l|rr|rr|rr} %
\toprule
        & \multicolumn{2}{c|}{\bf LinkedIn} & \multicolumn{2}{c|}{\bf X (Twitter)} & \multicolumn{1}{c}{\bf GitHub} \\
        \textbf{Dataset} & \textbf{Users} & \textbf{Posts} & \textbf{Users} & \textbf{Posts} & \textbf{Users} \\
        \midrule
        \textbf{\datasetPopular} & 100 & 985 & 100 & 11,998 & 100 \\
        \textbf{\datasetMalware} & 47 & 276 & 129 & 9,082 & 170 \\
        \textbf{\datasetLarge} & 0 & 0 & 878 & 97,369 & 878 \\
        \midrule
        \textbf{Total} & 147 & 1,261 & 1,107 & 118,449 & 1,148 \\
        \bottomrule
    \end{tabular}
\caption{Summary of our datasets. Each GitHub user has their profile bio as their single post.}
    \label{table:dataset}
\end{table}

We explain how we collect the dataset that we use in our study.
We apply \algo to a set of 1.2K users by starting from GitHub users and identifying self-reported social handles to LinkedIn and X (formerly Twitter). 
We provide a summary of our datasets in Table~\ref{table:dataset}.
The reported activity, namely the user posts, happens to span a time frame of 14 years (2010--2024).

\descr{Sourcing from GitHub.} To find tech users with active accounts on multiple platforms, we use the social links added by the users themselves to their GitHub profiles. %
More precisely, we start by crawling every public GitHub profile, saving the profile page of all 625K {\em popular} users, i.e., \popularDef. %
The most popular social platforms linked on these profiles include X and LinkedIn, with, respectively, 82K and 20K users providing links to them, followed by Instagram and YouTube, with 4.1K and 3.1K links. %
Thus, we opt to focus on X and LinkedIn in our cross-platform analysis.

\descr{The \datasetPopular, \datasetLarge, \datasetMalware datasets.} Out of the 625K popular GitHub authors, 14.7K have two or more social media links on their profile page, with 6.5K of them providing links to both X and LinkedIn.
We sample 100 users and collect their content.
We label this dataset as \datasetPopular.
Next, we sample 1K users from the 82K GitHub authors with links to the most common platform, X.
We exclude 122 users that have no text content on either platform, resulting in 878 users.
We label this dataset as \datasetLarge.
Finally, we build the \datasetMalware dataset with the 170 GitHub authors that have either a linked LinkedIn account, a linked X account, or both, and that host at least one malware repository with malicious intent~\cite{tania2024creating}.
From these 170 malware GitHub authors, we find 47 with valid LinkedIn accounts, 129 with valid X accounts, and 6 authors with both.

\descr{Data Collection.} For LinkedIn, we collect each of the users' posts on page one of their activity feed, denoted as ``All Activity'' on the main profile page, and the profile bio.
For X, we get all or up to the 200 most recent Tweets.
For GitHub, we collect the bio of each user from their profile page.

\subsection{Results}
\label{sec:casestudy}

We apply \algo on the datasets above to assess the relevance of cross-platform user profiling.

We revisit our choice of datasets.
As we discussed, we start from a set GitHub users, 
and our analysis focuses on their displayed personas on LinkedIn and X.
The combination of X and LinkedIn is fortuitous.
Arguably, X is typically associated with general social use, where users post about their activities, current events, news, and what they had for breakfast, among other things. By contrast, LinkedIn is a professional-oriented networking platform.
Therefore, it is plausible that users may selectively display different aspects of their persona and behave differently.
In other words, one might suppose that  LinkedIn posts are crafted to enhance the professional image of the user, while on X, users may be less curated and controlled. In fact, this may be especially true since Twitter became X, content moderation has been reduced and hate speech has been more prevalent~\cite{hickey2023auditing, benton2022hate}.

\begin{table}[t]
\centering
\setlength{\tabcolsep}{7pt}
\begin{tabular}{l|rr|r}
\toprule
& \multicolumn{2}{c|}{\bf \datasetPopular} & {\bf \datasetLarge}\\[1ex]
 & {\bf LinkedIn} & {\bf X}  & {\bf X} \\ \midrule %
Openness & 4.20 & 4.03 & 3.86 \\
Conscientious & 4.33 & 3.89 & 3.72\\ %
Extraversion & 3.63 & 3.98 & 3.72 \\ %
Agreeable & 4.08 & 4.08  & 3.90\\ %
Neuroticism & 1.79 & 2.91 & 2.89 \\ \bottomrule %
\end{tabular}
\caption{Averages OCEAN scores for \datasetPopular and \datasetLarge users.}
\label{table:platformstrengths}
\end{table}

\subsubsection{Single-Platform OCEAN Analysis.} Before diving into the cross-platform analysis, we first report, in Table~\ref{table:platformstrengths}, the average scores of the single-platform OCEAN personality traits for LinkedIn and X users in \datasetPopular as well as X users in \datasetLarge to gain insight into differences between the platforms as a whole.
Note how LinkedIn exhibits high scores for openness and conscientiousness, aligning with its use for professional networking.
Also, on X, extraversion and neuroticism get higher scores, which aligns with its social, less moderated nature.
\begin{figure}[t]
\centering
\includegraphics[width=\linewidth]{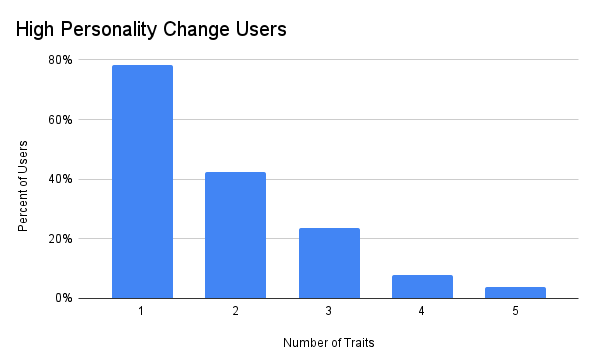}
    \caption{Magnitude of change in multiple OCEAN personality traits of users in \datasetPopular across LinkedIn and X.}
    \label{fig:high_change_users}
\end{figure}

\begin{table}[t]
\centering
    		\setlength{\tabcolsep}{6pt}
\begin{tabular}{rcrrr}
\toprule
&& \multicolumn{3}{c}{\bf Changed by}\\
 \textbf{\#Traits} && \textbf{1.0+} & \textbf{2.0+} & \textbf{3.0+} \\
\midrule
\textbf{1} && 78.2\% & 25.7\% & 3.0\% \\
\textbf{2} && 41.6\% & 6.9\% & 0\% \\
\textbf{3} && 23.8\% & 3.0\% & 0\% \\
\textbf{4} && 7.9\% & 0\% & 0\% \\
\textbf{5} && 4.0\% & 0\% & 0\% \\
\bottomrule
    \end{tabular}
\caption{Percentage of \datasetPopular users with changes in scores of more than 1.0, 2.0, and 3.0 on an increasing number of personality traits.}
    \label{table:highChangeUsersTable}
\end{table}

\subsubsection{Change in Displayed Personalities.}
Next, we set out to measure changes in displayed personalities across platforms.
As mentioned earlier, we are interested in noteworthy changes whereby an OCEAN score of a user (which lies in [1,5]) changes by at least one point in absolute terms. %

Overall, we find that most users in \datasetPopular exhibit consistent and noteworthy changes in the personality displayed across platforms.
Indeed, as shown in Figure~\ref{fig:high_change_users}, an overwhelming majority (77\%) of users change by at least one point in at least one personality trait.
Furthermore, almost 1 in 4 (24\%) users change by at least one point in at least three separate traits, with 4\% of users change by at least two points in all five traits.
A more detailed breakdown of the number of traits exhibiting change and the magnitude of the change is reported in Table~\ref{table:highChangeUsersTable}.

The personality changes are best exemplified by observing the overall change in neuroticism across platforms.
We find that 90\% of the users in \datasetPopular score below 2.5 neuroticism on LinkedIn, while only 40\% of users do so below 2.5 on X.
In fact, 95\% of the users increase in neuroticism from their LinkedIn to their X personas, with an average increase of 1.2 points per user.
We dig deeper by looking at the top and bottom 10\% users in \datasetPopular that exhibit, respectively, the most and the least change in neuroticism from LinkedIn to X. %
While the bottom 10\% users are more neurotic on LinkedIn (up to 1.3 points more on LinkedIn than X), the change is much more noticeable for the top 10\% of users (up to 3.1 points more neurotic on X).

We observe similar results on the \datasetMalware dataset.
As discussed in Sec.~\ref{sec:dataset}, these users are sampled from a very different population, yet they show the same trends with respect to changing their displayed personality across LinkedIn and X.
For instance, their average neuroticism on LinkedIn is 1.7, while on X, it is 2.6.
Also, while only 41\% of \datasetMalware users have neuroticism above 2.0 on LinkedIn, 76\% do so on X.

We also find significant {\em directional} changes in OCEAN traits across platforms, with \datasetPopular users displaying 1.1 higher neuroticism and 0.4 lower conscientiousness on X than on LinkedIn, on average.
Two-sample Kolmogorov–Smirnov (KS) tests~\cite{smirnov1939estimation} %
confirm statistically significant differences across LinkedIn and X for the CDFs of both neuroticism and conscientiousness with, respectively, $D=0.63$ ($p < 0.001$) and $D=0.46$ ($p< 0.001$).
We find smaller directional changes for extraversion of 0.36 higher on X with a KS statistic of $D=0.31$ ($p < 0.001$), and for openness of 0.17 lower on X with a KS statistic of $D=0.30$ ($p < 0.001$).
Agreeableness is largely consistent across platforms, with an average increase of 0.01 from LinkedIn to X, and a KS statistic of $D=0.20$ ($p < 0.05$).

In summary, %
we observe that users in our dataset tend to become more ``unruly'' but also more outgoing in their activity on X, supported by increased neuroticism and 
extraversion, as opposed to their activity on LinkedIn.

\begin{table}[t]
\centering
\begin{tabular}{lr|lr}
\toprule
    \multicolumn{2}{c|}{\bf LinkedIn} & \multicolumn{2}{c}{\bf X}\\
    {\bf Interest} & {\bf RFreq} & {\bf Interest} & {\bf RFreq} \\
    \midrule
    Web Development & 14.7\% & Development Tools & 15.2\% \\
    Development Tools & 12.6\% & Script Languages & 10.3\% \\
    Script Languages & 8.0\% & Web Development & 8.5\% \\
    Open Source & 7.6\% & Java & 3.6\% \\
    Java & 2.5\% & Business Software & 3.0\% \\
\bottomrule
\end{tabular}
\caption{Relative frequency (RFreq) of the top five most common leaf-level professional interests for \datasetPopular users on LinkedIn and X.}%
\label{table:professional_interests_popular_linkedin}
\end{table}

\subsubsection{Professional Interests.}
Next, we turn to professional interests, aiming to assess whether tech users reveal their professional interests equally across their active platforms.
We do so with two levels of granularity: high-level professional interest categories and leaf-level categories (see Section~\ref{sec:profiling}).
Given that the users in our dataset originate from GitHub, we designate the high-level categories of `Computers \& Electronics' and `Internet \& Telecom,' along with their respective leaf-level categories, as professional interest categories, and the rest as personal interest categories.
For the high-level categories, we observe no significant differences between LinkedIn and X.
For \datasetPopular, 81.5\% and 86.7\% of the categories are under `Computers \& Electronics' while 18.5\% and 13.3\% are under `Internet \& Telecom,' respectively, for LinkedIn and X. 
This is unsurprising given that these are popular tech users we discover from GitHub, who likely discuss their work on the platforms linked from their GitHub profiles.
In fact, we get very similar results for the high-level professional categories for \datasetMalware.
For \datasetLarge, we find nearly an identical category distribution of 85.8\% and 14.2\% between `Computers \& Electronics' and `Internet \& Telecom,' respectively, on X.

The same trend, although to a lesser extent, remains for professional categories at the leaf-level granularity.
In \datasetPopular, four leaf-level professional categories appear in the top five categories for both LinkedIn and X (namely, `Web Design \& Development,' `Development Tools,' `Scripting Languages,' and `Java'), as reported in Table~\ref{table:professional_interests_popular_linkedin}.
In \datasetMalware, we again find four categories in common among the top five on each platform: `Network Security, `Development Tools,' `Hacking \& Cracking,' and `Scripting Languages.'
In \datasetLarge, the top five categories for X are `Web Design \& Development,' `Development Tools,' `Open Source,' `Database Management', and `Operating Systems', aligning with the top categories in the other datasets.%

\subsubsection{Personal Interests.}
A complementary aspect to extracting users' professional interests is determining their personal interests to allow for more holistic profiling of users.
On this front, we find a significant difference across LinkedIn and X. %
In terms of high-level personal interests, we find common behavior across popular and malware tech users.
On LinkedIn, the three most frequent high-level categories are `Jobs \& Education,' `Science,' and `Online Communities,' accounting for 44.9\%, 18.2\%, and 10.2\% of categories, respectively, for \datasetPopular and 48.4\%, 9.9\%, and 8.8\%, respectively, for \datasetMalware.
On X, both popular and malware tech users display the most interest in `Online Communities,' with a total of 50.9\% and 46.3\% of categories, respectively.
However, their less common interests diverge, with \datasetPopular having `Science,' `Reference,' and `Jobs \& Education,' with \datasetMalware having `Arts \& Entertainment,' `Games,' and `News' as their next three most frequent categories, respectively.
The interests of \datasetLarge lie in between the other datasets, having `Jobs \& Education,' `Science,' `Online Communities,' and `Arts \& Entertainment' as their most frequent categories.

Investigating the leaf-level personal interest categories reveals a similar pattern.
On LinkedIn, four of the five most frequent leaf-level personal interests for \datasetPopular and \datasetMalware are the same.
The one category not in common is `How-To, DIY, \& Expert Content' for \datasetPopular versus `Colleges \& Universities' for \datasetMalware.
On X, the two most frequent categories are identical across all three datasets (\datasetPopular, \datasetMalware, and \datasetLarge), and the less frequent interests diverge.

Looking at the most frequent leaf-level interests across LinkedIn and X reveals only one category in common in each dataset, i.e., `Machine Learning \& Artificial Intelligence' for \datasetPopular and `Social Networks' for \datasetMalware.
In summary, we find that tech users display different personal interests on different platforms and that both popular and malware tech users display similar personal interests on the same platform.

\begin{table}[t]
\centering
\begin{tabular}{l|rrr}
\toprule
\textbf{Trait} & \textbf{Correlation} & \textbf{P-value} \\
\midrule
\textbf{Openness} & 0.055 & 0.542 \\
\textbf{Conscientiousness} & -0.023 & 0.798 \\
\textbf{Extraversion} & 0.239 & 0.007 \\
\textbf{Agreeableness} & -0.120 & 0.179 \\
\textbf{Neuroticism} & 0.373 & 0.000 \\
\bottomrule
    \end{tabular}
\caption{Correlation between offensive behavior and OCEAN traits for \datasetMalware users on X.}
   
    \label{table:offensiveCorrelation}
\end{table}

\subsubsection{Offensive Behavior.}
Finally, we compare offensive behavior across platforms.
As expected, this is more common on X than on LinkedIn, as shown in Table~\ref{table:platformstrengths} for \datasetPopular users.
We find that tech users show no signs of offensive behavior on LinkedIn, nor in any of their profile bios on GitHub, but 10\% of these same users act in an offensive manner on X.
As the users in \datasetPopular are a sample of popular GitHub developers, they likely use LinkedIn to form connections to improve their professional prominence by advertising their work on GitHub, which can leave X as their informal venting platform.
For example, we find a user (denoted as {\tt xyz} to protect their anonymity) that writes well-structured LinkedIn posts about their Medium blog where they discuss a ChatBot application they published on GitHub, while on X they can be seen replying to users with insults of various kinds. %

We also measure the correlation between certain OCEAN traits and offensiveness on X using Point Biserial Correlation~\cite{kornbrot2014point}.
As reported in Table~\ref{table:offensiveCorrelation}, we find positive correlations of 0.24 (p-value of 0.01) for extraversion and 0.37 (p-value of 0.00) between neuroticism and offensive behavior for \datasetMalware users.
For \datasetPopular and \datasetLarge, the correlation is almost identical (0.25, with p-value of 0.01, and 0.22, with p-value of 0.00, respectively). %
This suggests that the neurotic users in our dataset are moderately likely to display offensive behavior, although, interestingly, only on X and not on LinkedIn or GitHub.
For instance, the offensive user {\tt xyz} displays a neuroticism of 3.9 on X. %

\subsection{User Groups}
Finally, we set out to find interesting groups of tech users with similar displayed personalities or similar changes in personalities across platforms.
This provides us with insight into common profiles and behaviors.
As mentioned earlier, the grouping can focus on any combination of features, on single or multiple platforms; next, we explore some possible choices.

\begin{table}[t]
\centering
\begin{tabular}{l|rrrrr}
    \toprule
    & \textbf{Open} & \textbf{Cons} & \textbf{Extr} & \textbf{Agre} & \textbf{Neur} \\
    \midrule
    \textbf{Cluster 1} & 4.73 & 4.76 & 4.39 & 4.67 & 1.85 \\ %
    \textbf{Cluster 2} & 4.03 & 3.98 & 3.57 & 3.98 & 1.93 \\ %
    \textbf{Cluster 3} & 3.73 & 4.50 & 3.13 & 4.55 & 1.18 \\ %
    \textbf{Cluster 4} & 4.41 & 4.36 & 2.32 & 3.02 & 1.06 \\ %
    \textbf{Cluster 5} & 2.95 & 3.81 & 2.45 & 2.58 & 1.00 \\ %
    \textbf{Cluster 6} & 4.68 & 4.45 & 4.23 & 4.47 & 3.55 \\ %
    \textbf{Cluster 7} & 1.80 & 2.20 & 1.60 & 2.00 & 3.50 \\ %
    \bottomrule
\end{tabular}
\caption{Average OCEAN traits of the personality clusters of \datasetPopular users on LinkedIn.}
\label{table:linkedinClusters}
\end{table}

\begin{table}[t]
\centering
\begin{tabular}{l|rrrrr}
    \toprule
    & \textbf{Open} & \textbf{Cons} & \textbf{Extr} & \textbf{Agre} & \textbf{Neur} \\
    \midrule
    \textbf{Cluster 1} & 4.11 & 4.00 & 4.09 & 4.16 & 2.97 \\ %
    \textbf{Cluster 2} & 3.33 & 2.94 & 2.95 & 3.35 & 2.20 \\ %
    \bottomrule
\end{tabular}
\caption{Average OCEAN traits of the two personality clusters of \datasetPopular users on X. Cluster 1 captures users with higher OCEAN traits, while Cluster 2 those with lower traits.}
\label{table:twitterClusters}
\end{table}

\subsubsection{Personality Groups.}
We cluster tech users by their personality as a five-dimensional vector of their OCEAN trait values.
As discussed in Section~\ref{sec:integrate}, we use the $k$-means silhouette score method to determine the optimal $k$.
For \datasetPopular, we find that $k=7$ on LinkedIn and $k=2$ on X.
We report a summary of the LinkedIn clusters in Table~\ref{table:linkedinClusters} and X clusters in Table~\ref{table:twitterClusters}.
While it may seem intuitive that the same users on LinkedIn would belong to the same cluster on X, %
that is not the case, as we find that six out of the seven LinkedIn clusters have membership in both the two X clusters.
This is corroborated by users on X being grouped into clusters of distinct ``low'' and ``high'' OCEAN trait values, while the clusters on Linkedin do not have such clear distinctions.
To better illustrate this phenomenon, let us consider the example of user {\tt abc} (again, identity hidden for anonymity)
from \datasetPopular.
They are part of Cluster 2 on LinkedIn (cf.~Table~\ref{table:linkedinClusters}), characterized by medium to high values in all personality traits except for neuroticism, and of Cluster 2 on X (cf.~Table~\ref{table:twitterClusters}), which, on average, has higher neuroticism and lower values in all other personality traits.
Upon closer inspection, we notice {\tt abc} often creates long, detailed hiring posts for their company on LinkedIn and advertising posts for their online machine learning course.
Whereas on X, they frequently post short statements on the current stock market trends and advocate for political activism causes.
This behavior aligns with their displayed personality on each platform, with respective OCEAN trait values of 4.0, 4.1, 3.1, 4.1, 2.6 on LinkedIn and 1.8, 1.6, 1.9, 2.9, and 3.2 on X.
Overall, this suggests that users in our dataset may display drastically different personalities across platforms due to entirely different usage purposes for each platform.

\subsubsection{Groups of Tech Users with Significant Cross-Platform Changes.}
Next, we use $k$-means to identify clusters of \datasetPopular users that display similar absolute change in personality across LinkedIn and X.
We determine that the optimal number of clusters is $k=2$ using the same silhouette score method.
The resulting two clusters distinguish between high-change tech users, with average personality changes of 1.06, 1.11, 1.52, 1.16, 1.46, respective to each OCEAN trait, and low-change tech users with an average change of 0.32, 0.45, 0.47, 0.42, 1.07, respectively.
Interestingly, in both clusters, the average change in neuroticism is relatively high -- over one point -- indicating that these users typically exhibit different neuroticism across platforms irrespective of their other personality traits.
Identifying groups of users with similar changes in personality could allow for predicting how a user will interact on a different platform by observing how other users in that cluster interact on the platform.

\section{Related Work}
\label{sec:relatedwork}

Overall, most previous work focuses on user profiling on a single platform or does not address the problem as framed here.
We can place related work into the following groups.

\descr{Single-platform user profiling.}
Many  studies profile the personalities of users on a single platform, which can be seen as a subset of the work we do.
They do so by combining multiple pre-trained language models for more accurate predictions~\cite{christian2021text}, utilizing word embeddings from public tweets in a supervised learning approach~\cite{carducci2018twitpersonality}, and leveraging follower, following, and listed counts on profiles~\cite{quercia2011our}.
Other works correlate users' linguistic features with traits from the Myer-Briggs Type Indicator (MBTI) personality model~\cite{wang2015understanding}, or correlate social media use with personality traits, demographic attributes, and general life outlook~\cite{ozguven2013relationship}.
Researchers have also studied how different personalities may be associated with different motives for using social media~\cite{chen2023people}. %
Finally, recent efforts rely on LLMs to infer online users' OCEAN personality traits, yielding similar results to supervised models~\cite{peters2024large}.

\descr{User disambiguation methods.}
Some studies consider two platforms, but focus on user disambiguation, namely matching user identities across platforms ~\cite{goga2015reliability}.
These efforts can be seen as an enabling capability for our work  as they can provide more data in the Matching stage of \algo.
These efforts include using stylometry to match users from their written content\cite{vosoughi2015digital}, linking user generated content to the users it originated from \cite{li2017comparison} \cite{alonso2021writer}, or utilizing metadata~\cite{malhotra2012studying} and linguistic features~\cite{userIdentityLinkage2020} to match users across platforms.
One effort leverages the structure of online usernames to identify user matches~\cite{masud2023geekman}.
Other efforts leverage user-posted URLs~\cite{treves2023rurlman} and time features from user activity~\cite{johansson2015timeprints} to link users across platforms.

\descr{Targeted and niche cross-platform studies.}
Several studies involve different platforms but with a different focus than our work.
For example, many studies answer niche and targeted questions, such as correlating personality with a certain behavior~\cite{samani2018cross}, identifying preferences, such as the user's taste in music ~\cite{gu2018cross}, or improving user modeling in recommender systems~\cite{abel2013cross}.
Other related studies use the overlap of a small number of users between two social platforms to predict the behavior of non-overlapped users~\cite{jiang2016little}, or identify differences in linguistic characteristics of users' text structures on different platforms~\cite{liu2022cross}. %
One study demonstrates that users often use different types of profile pictures and descriptions on different social platforms~\cite{zhong2017wearing}, and another determines that personality traits help explain the types of social media people use~\cite{vaid2021uses}.

\section{Discussion \& Conclusion}
\label{sec:discussion}
This paper presented \algo, a systematic methodology for extracting and comparing users' displayed personas across platforms.
\algo handles multiple types of platforms and extracts multiple types of information for each user. %
Its main module infers OCEAN personality traits~\cite{goldberg1990alternative} 
on a five-point scale, achieving a reasonably low average Root Mean Squared Error (RMSE) of 1.1 over a labeled dataset~\cite{celli2013workshop}.

We build a dataset of 1.2K tech users sourced from GitHub who self-report their LinkedIn and/or X accounts and apply \algo to it to shed light on their displayed personality changes across different platforms.
For instance, we show that
the overwhelming majority of our users change by at least one point %
in at least one personality trait, with neuroticism exhibiting the most differences between LinkedIn and X. 
Overall, we are confident \algo can support more nuanced profiling of users, relying on multiple platforms and types of information.
\descr{Limitations.}
Naturally, our work is not without limitations. 
We do not fine-tune or train models specifically to predict users' personalities -- e.g., %
using ad-hoc language analysis tools -- since, even untuned, Gemini provides reasonable accuracy and requires little overhead to implement, maximizing practicality of use.
Also, we are inherently unable to use content that a user has deleted and can only infer displayed personality based on text from public posts, which may paint an incomplete picture; however, the inferred personality can still be used as a meaningful comparison. %

Moreover, any  measurement-driven study has to
defend its choice of data with respect to its generalizability of its results.
First, we want to argue that the purpose of the study is to show that:
(a) our approach works and
(b) it has the potential to provide interesting insights.
In that regard, the specific dataset or the exact magnitude of the observed differences is secondary.
Having said that, our datasets are sourced from profiles linked by GitHub users, thus adding a technology-oriented bias.
Our analysis of tech users' personal and professional interests shows relatively little change across platforms, potentially as a result of this bias.
At the same time, this also further highlights the importance of cross-platform profiling, since users may post about the same topics, but they may do so in different manner across different platforms, as attested to by the widespread per-user displayed personality changes.
Finally, \datasetLarge does not contain LinkedIn data; thus, its analysis is limited to making conclusions about user personas on X.
Nonetheless, we take advantage of its relatively larger size to confirm conclusions derived from the smaller \datasetPopular and \datasetMalware, which we plan to expand in future work.

\descr{Ethical Considerations and Design Implications.}
We follow best practices according to the ACM-community code of ethics and professional conduct~\cite{gotterbarn2018acm}.
All the data collected and analyzed as part of this study is publicly available and voluntarily posted by the users.
Furthermore, our analysis primarily focuses on aggregate statistics.
To ensure reproducibility and comply with the FAIR data principles~\cite{fair}, we will share the source code of our implementations as well as our datasets with researchers upon request, with datasets limited to user IDs rather than raw data. %
Tools that facilitate the mining of user data for the purpose of psychological profiling can, understandably, be seen as privacy-invasive.
Note, however, that we only study users who self-select to link their profiles across platforms -- more precisely, users who report their X and/or Linkedin profiles from the GitHub page.

Due consideration must also be given to the potential weaponization of cross-platforming personality profiling, e.g., for accurate micro-targeting of political messages~\cite{berghel2018malice}, ads, disinformation, hate campaigns, and so on~\cite{farkas2018disguised,hristakieva2022spread}. %
However, refraining from pursuing this kind of research would not make these concerns go away, as profiling is already routinely done in the wild, at an immense scale, by social network providers and data mining/data broker companies~\cite{aiello2020social}.
Rather, our work can be seen as a warning for users, practitioners, and policymakers as it displays what kind of knowledge and how accurately companies and organizations can extract from publicly available information the users voluntarily reveal, especially when multiple platforms can be combined.

Moreover, potential inaccuracy in profiling may lead to inaccurate inferences about users' personas. 
However, we do not use \algo at single-user granularity but, rather, at an aggregate level to elicit trends about cross-platform changes.

\descr{Extensions to \algo.}
In future work, we plan to add an automatic translation feature due to the high pervasiveness of foreign language content. %
This may potentially lead to more accurate profiling, e.g., in terms of interest extraction.
However, due to the nuance in language possibly being lost in translation, e.g., for language-specific offensive behavior, we will likely use both translated and original content. %

In addition, we will expand on the personality inference module by implementing procedures for existing methods, such as incorporating linguistic features from user-generated text via LIWC~\cite{pennebaker2007linguistic}, as well as by including multi-modal data, such as images or videos.
Finally, we plan to extend our evaluation to more platforms; in our sample of popular GitHub users, we also find a large number of users linking to their Mastodon, YouTube, or Telegram accounts, as well as personal blogs, which may offer a larger variety of platforms for a more extensive study.

\balance
\bibliographystyle{ACM-Reference-Format}

\end{document}